\documentclass[pre,aps,twocolumn,
pdflatex,
superscriptaddress,floats,showpacs,10pt]{revtex4-1}
\usepackage{graphicx}
\usepackage{dcolumn}
\usepackage{bm}
\usepackage{amssymb}
\usepackage{amsmath}
\usepackage{textcomp}
\usepackage{color}

\begin{document}

\title{Explosion of relativistic electron vortices in laser plasmas}

\author{K. V. Lezhnin}
\email{lezhnin@phystech.edu}
\affiliation{Institute of Physics of the ASCR, ELI-Beamlines, Na Slovance 2, 18221 Prague, Czech Republic}
\affiliation{Moscow Institute of Physics and Technology, 
Institutskiy per. 9, Dolgoprudny, Moscow Region 141700, Russia}

\author{F. F. Kamenets}
\affiliation{Moscow Institute of Physics and Technology, 
Institutskiy per. 9, Dolgoprudny, Moscow Region 141700, Russia}

\author{T. Zh. Esirkepov}
\affiliation{National Institutes for Quantum and Radiological Sciences and Technology, 8-1-7 Umemidai, Kizugawa, Kyoto 619-0215, Japan}

\author{S. V. Bulanov}
\affiliation{National Institutes for Quantum and Radiological Sciences and Technology, 8-1-7 Umemidai, Kizugawa, Kyoto 619-0215, Japan}
\affiliation{A. M. Prokhorov General Phys. Inst. of RAS, Vavilov Str. 38
, Moscow 119991, Russia}

\author{Y. Gu}
\affiliation{Institute of Physics of the ASCR, ELI-Beamlines, Na Slovance 2, 18221 Prague, Czech Republic}
\author{S. Weber}
\affiliation{Institute of Physics of the ASCR, ELI-Beamlines, Na Slovance 2, 18221 Prague, Czech Republic}
\author{G. Korn}
\affiliation{Institute of Physics of the ASCR, ELI-Beamlines, Na Slovance 2, 18221 Prague, Czech Republic}

\date{\today}

\begin{abstract}
The  interaction of high intensity laser radiation with underdense plasma may lead to the formation of electron vortices. Though being quasistationary on an electron timescales, these structures tend to expand on a proton timescale due to Coloumb repulsion of ions. Using a simple analytical model of a stationary vortex as initial condition, 2D PIC simulations are performed. A number of effects are observed such as vortex boundary field intensification,
multistream instabilities at the vortex boundary, and bending of the vortex boundary with the subsequent transformation into smaller electron vortices.

\bigskip

\noindent Keywords: 
Relativistic laser plasmas, 
Electron vortices, 
Quasi-static magnetic field,
Particle-in-Cell simulation
\end{abstract}

\pacs{52.38.Kd, 52.65.Rr}
\maketitle

\section{Introduction}

Formation of localized coherent structures during the interaction of 
intense laser pulses with plasmas is an important topic of the laser plasmas research, which is vital for diagnostic purposes in the experiments with laser ion acceleration, the fast ignition of controlled thermonuclear fusion, the investigation of warm dense matter, high energy density phenomena,
and laboratory astrophysics
(see article \cite{VORTEXEXP, SOLITON1} and references therein). 
When the laser pulse interacts with a homogeneous plasma region, we expect the pulse to penetrate inside the plasma and propogate with minor energy losses in case of underdense plasma. Quantitavely speaking, we expect laser pulse to penetrate if $\omega_0 > \omega_{\rm pe}$, where $\omega_0$ is the laser carrier frequency and $\omega_{\rm pe}=\sqrt{4 \pi n_e e^2 /m_e}$ is the electron plasma frequency. The larger the $\omega_0/\omega_{\rm pe}$ ratio is, the lower is the laser pulse depletion rate. However, eventually a finite duration laser pulse completely depletes due to stimulated Raman scattering, various pulse filamentation instabilities and transformation to various localized coherent structures \cite{REVPLASMPHYS}. These processes lead to the formation of Langmuir waves, electromagnetic solitons, and electron vortices retaining electromagnetic energy.

Besides self-focusing channels and Langmuir waves, we consider two classes of coherent structures that happen to form in laser plasmas simulations. In two-dimentional plasmas, we can distinguish solitons \cite{SOLITON1, SOLITON2} and electron vortices \cite{VORTEX}.
Electron vortices are coherent localized structures, which keep quasistatic magnetic flux locked inside the electron cavity. Solitons enclose the electromagnetic wave inside the density gap. Though having slightly different conditions of generation, in 3D case these structures are usually combined in one, posessing some features from both of them \cite{CENTAUR}.
On an electron timescale $\omega_{\rm pe}^{-1}$, these coherent structures are quasistationary. In the case of immobile ions, an analytical stationary solution exists, see Section II. However, on a proton timescale $\omega_{\rm pi}^{-1}=\sqrt{m_p/m_e}~\omega_{\rm pe}^{-1}$, we expect these vortices to evolve, as the uncompensated proton charge results in Coloumb explosion of the whole vortex \cite{Gordeev}. Hereafter, we will call it ``vortex explosion''.

Here we investigate the structure and evolution of the relativistic electron vortices. We present a simple analytical vortex model which illustrates the main features of realistic vortex structure. We carry out two-dimensional (2D) Particle-in-Cell (PIC) simulations
using the REMP code
based on the density decomposition scheme \cite{REMP}.
We discuss in detail a number of effects occurring during the evolution of electron vortices: 
the Coulomb explosion of the uncompensated proton core of the vortex, multistream processes that lead to shell-like structure formation, instability at the vortex boundary leading to the edge field intensification and transformation into smaller vortices.
The paper is organized as follows.
In the next Section, the analytical solution for stationary vortex structure is presented.
In Section III, we describe the simulation parameters.
In Section IV, we discuss simulation results of the evolution of the electron vortex, addressing the observed effects.
In the concluding Section, we summarize the results obtained.

\section{Stationary electron vortex structure}

Before discussion of the simulations of the vortex structure evolution, we present theoretical estimates for the analytical structure of the electron vortex in the magnetized collisionless plasma region.

\subsection{General equations}

We start from Maxwell equations
\begin{eqnarray}
\partial_t {\bf E} = \nabla\times{\bf B}-{\bf J}
\, , \label{eq:Et}\\
\partial_t{\bf B}= -\nabla\times{\bf E} 
\, , \\
\nabla\cdot{\bf E}= \rho 
\, , \\
\nabla\cdot{\bf B}= 0 
\, , \label{eq:divB}
\end{eqnarray}
and relativistic plasma motion equations
\begin{eqnarray}
\partial_t (\gamma_e {\bf v}_e) + ({\bf v}_e\cdot\nabla)(\gamma_e {\bf v}_e) = -2\pi ({\bf E}+{\bf v}_e\times {\bf B})
\, , \\
\partial_t (\gamma_i {\bf v}_i) + ({\bf v}_i\cdot\nabla)(\gamma_i {\bf v}_i) = 2\pi Z({\bf E}+{\bf v}_i\times {\bf B})
\, . \label{eq:vit}
\end{eqnarray}
Here, current ${\bf J}$ and charge density $\rho$ can be written as
\begin{eqnarray}
{\bf J} = Z n_i {\bf v}_i - n_e {\bf v}_e
\, , \label{eq:J}\\
\rho = Z n_i - n_e
\, , \label{eq:n}
\end{eqnarray}
and relativistic gamma-factors for electrons and ions are
\begin{eqnarray}
\gamma_e = (1-v_e^2)^{-1/2}
\, , \label{eq:ge}\\
\gamma_i = (1-v_i^2)^{-1/2}
\, . \label{eq:gi}
\end{eqnarray}

\noindent
All equations are in dimensionless form. Hereafter, we will measure spatial parameters in $\lambda= 1 \mu m$, temporal -- in $2 \pi / \omega_0 = \lambda / c$, densities -- in critical densities $n_{\rm cr}=m_e \omega_0 / 4 \pi e^2$, electromagnetic fields -- in $E_0 = m_e \omega_0 c / e$, where $m_e$ is electron mass, $e$ is the absolute value of electron charge, $c$ is the speed of light in vacuum. In addition, we note that in dimentionless units $e/m$ becomes $2\pi$ and velocities are measured in fractions of light speed in vacuum.

\subsection{2D cylindrical configuration}
          
Changing to cylindrical coordiantes $(r,\theta,z)$,
we further assume asymmetry and homogeneity along the $z$-axis, i.e.
$\partial_\theta = \partial_z = 0$.
We denote vector components by subscripts corresponding to respective coordinates, e.g., for electric field E components are $E_r, E_\theta, E_z$.
Besides, we assume that
\begin{eqnarray}
E_z=0
\, , \\
B_r = 0 \, , B_\theta = 0
\, , \\
v_{ez} = 0 \, , v_{iz} = 0
\, .
\end{eqnarray}

Then Eq. (\ref{eq:divB}) becomes identity and
from 
Eqs. (\ref{eq:Et})-(\ref{eq:vit}) together with definitions
Eqs. (\ref{eq:J})-(\ref{eq:n}) we obtain
\begin{gather}
\partial_t E_r = n_e v_{er} - Z n_i v_{ir} \,, \label{eq:Ert-cyl} \\
\partial_t E_\theta = -\partial_r B_z + n_e v_{e\theta} - Z n_i v_{i\theta}
\, , \label{eq:Ett-cyl}\\
\partial_t B_z = - \frac{\partial_r(r E_\theta)}{r}
\, , \label{eq:Bzt-cyl}\\
\frac{\partial_r(r E_r)}{r} = Z n_i - n_e
\, , \label{eq:Err-cyl}\\
\partial_t (\gamma_e v_{er}) + v_{er}\partial_r (\gamma_e v_{er}) 
- \frac{\gamma_e v_{e\theta}^2}{r} = -2\pi(E_r + v_{e\theta} B_z)
\, , \label{eq:vert-cyl}\\
\partial_t (\gamma_e v_{e\theta}) + v_{er}\partial_r  (\gamma_e v_{e\theta}) 
+ \frac{\gamma_e v_{er} v_{e\theta}}{r} = -2\pi(E_\theta - v_{er} B_z)
\, , \label{eq:vett-cyl}\\
\partial_t (\gamma_i v_{ir}) + v_{ir}\partial_r (\gamma_i v_{ir})
- \frac{\gamma_i v_{i\theta}^2}{r} = 2\pi Z(E_r + v_{i\theta} B_z)
\, , \label{eq:virt-cyl}\\
\partial_t (\gamma_i v_{i\theta}) + v_{ir}\partial_r (\gamma_i v_{i\theta}) 
+ \frac{\gamma_i v_{ir} v_{i\theta}}{r} = 2\pi Z(E_\theta - v_{ir} B_z)
\, .
\label{eq:vitt-cyl}
\end{gather}

%
We also assume that plasma is neutral at infinity
and the total charge is zero:
\begin{equation}
n_e=Zn_i {\rm\ for\ } r\rightarrow\infty
\, , \quad
\int(n_e - Zn_i)r dr = 0
\, . \label{eq:n-cyl}
\end{equation}

\subsection{2D stationary vortex in electron fluid}

Now we assume
$\partial_t = 0$, 
$E_\theta = 0$,
$v_{er} = 0$,
$Z n_i = Z n_{i0} = 1$,
${\bf v}_i = 0$
and neglect Eqs (\ref{eq:virt-cyl})-(\ref{eq:vitt-cyl}).
Then Eqs (\ref{eq:Ert-cyl}), (\ref{eq:Bzt-cyl}) and (\ref{eq:vett-cyl})
become identities
and from Eqs (\ref{eq:Ett-cyl}), (\ref{eq:Err-cyl}) and (\ref{eq:vert-cyl})
we obtain
\begin{eqnarray}
B_z^\prime = n_e v_{e\theta}
\, , \label{eq:Bzr}\\
E_r^\prime + \frac{E_r}{r} = 1 - n_e
\, , \label{eq:Err}\\
E_r = - v_{e\theta} B_z + \frac{v_{e\theta}^2}{2\pi r \sqrt{1-v_{e\theta}^2}}
\, , \label{eq:EBv}
\end{eqnarray}
where prime denotes derivative with respect to $r$,
$B_z^\prime = \partial_r B_z$.

\subsection{Non-relativistic limit}

Neglecting the last term in Eq (\ref{eq:EBv})
in the limit of non-relativistic electron motion, $|v_{e\theta}|\ll 1$,
and resolving this system wit respect to $v_{e\theta}$, $n_e$ and $B_z$
we obtain
\begin{eqnarray}
v_{e\theta} = -E_r/B_z
\, , \label{sol:vet}\\
n_e = 1 - E_r^\prime - \frac{E_r}{r}
\, , \label{sol:ne}\\
B_z^2(r) = B_{0}^2 +E_r^2(r) + 2\int\limits_0^r (E_r/r-1) E_r  dr
\, , \label{sol:Bz}
\end{eqnarray}
where we assume $E_r(r=0)=0$ and $B_z(r=0)=B_0>0$.
The system Eqs (\ref{sol:vet})-(\ref{sol:Bz})
gives the solution provided that a reasonable profile is set for $E_r(r)$. There could be infinite number of vortex shapes, however not all of them are stable.
Here we do not consider the problem of stability.

\begin{figure}[hb]
\begin{center}\includegraphics[width=8cm]{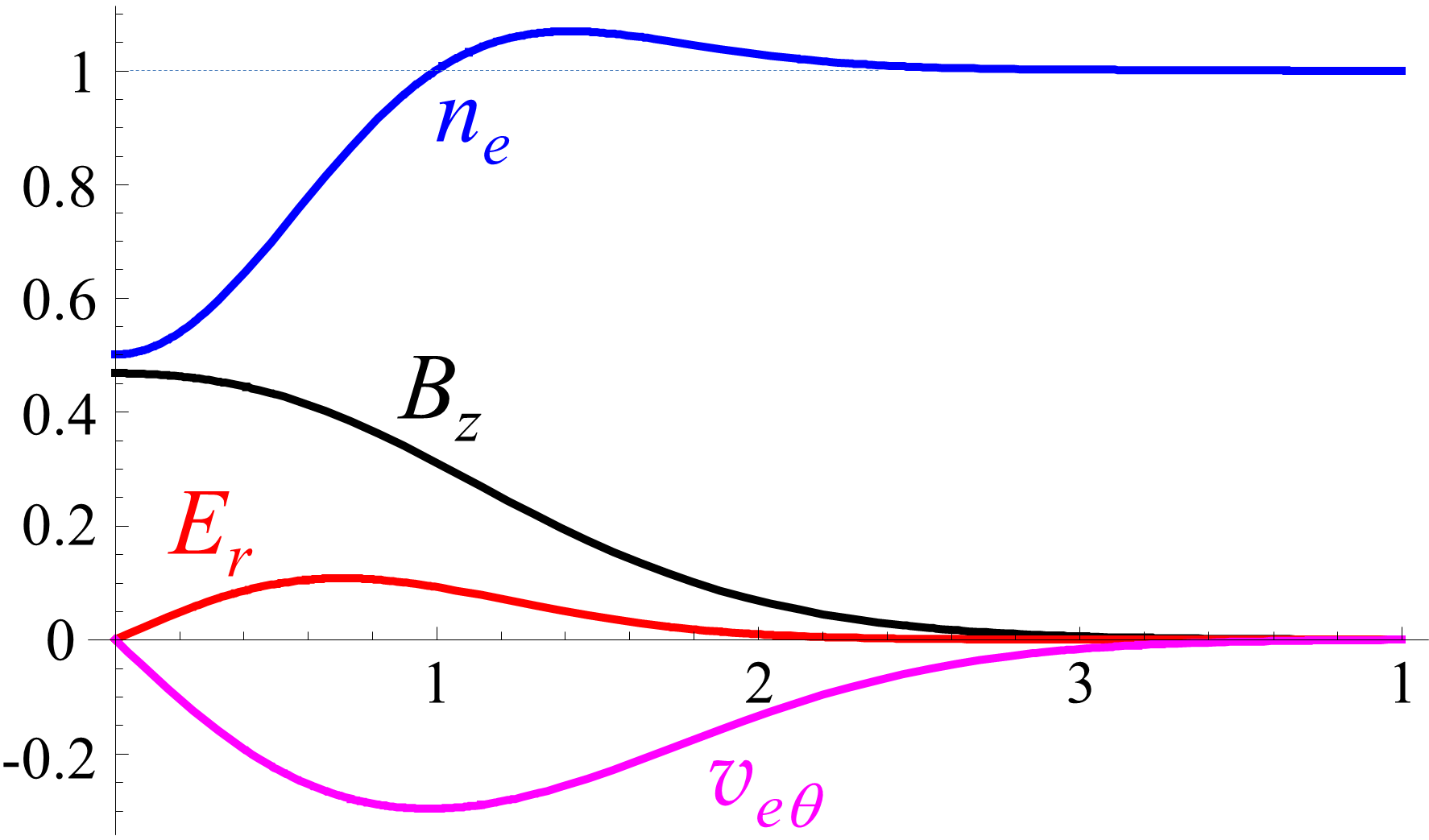}
\end{center}
\caption{\label{fig:1}
$E_r$, $B_z$, $n_e$ and $v_{e\theta}$ in the
vortex for $\alpha=1,E_0=1/4$.
}
\end{figure}

We assume the dependence of $E_r$ on $r$ as
\begin{equation}
E_r(r) = E_0 r e^{-\alpha r^2}.
\, \label{def:Er}
\end{equation}
It corresponds to a situation when the inner region of the vortex is represented by the homogeneous proton region, and to some extent in cyllindrical geometry $E_r$ grows lineary with radius, until the electric field is being shielded by outside plasmas.
Then from Eq (\ref{sol:Bz})
we obtain
\begin{equation}
\begin{split}
B_z^2(r) = \left(B_0^2 + \frac{E_0^2}{2\alpha} - \frac{E_0}{\alpha}\right)
+ \frac{E_0}{\alpha} e^{-\alpha r^2}
 \\ +E_0^2\left(r^2 - \frac{1}{2\alpha}\right) e^{-\alpha r^2}
 \label{def:Er}
\end{split}
\end{equation}
with the following asymptotic at $r\rightarrow 0$
\begin{equation}
B_z(r) = B_0 + \frac{E_0(2E_0-1)}{2B_0}r^2 + o(r^3)
\, .
\end{equation}
We assume that $B_z$ vanishes at infinity,
which leads to
\begin{equation}
B_0 = \sqrt{\frac{E_0}{2\alpha} (2-E_0)}
\, . \label{sol:B0}
\end{equation}
In addition, we impose the condition on $B_z$ 
to have a local maximum at $r=0$.
Then 
\begin{equation}
0<E_0<1/2
\, ,
\end{equation}
which ensures positive number under the root sign in Eq (\ref{sol:B0}).
For $\alpha=1,E_0=1/4$, see the profiles in Fig. \ref{fig:1}.
This simple vortex model highlights the main features of electron vortex - electron cavitation, localization of the magnetic flux inside the electron cavity and electron density increase in the vicinity of vortex boundary \citep{Gordeev, CAVITATION}. Though we are going to consider mildly relativistic electron vortices, the above consideration of nonrelativistic case fairly reproduces the main features we are going to observe in the next Section.


%
%
%
%
 
\section{Simulation configuration}

In order to investigate the processes that occur during 
the electron vortex evolution, we conduct a series of 2D PIC simulations. 

In our simulations, we set the homogeneous square of cold electron-proton plasma with the density about $n_e/n_{\rm cr} = 0.04$ and size of $38 \lambda \times 38 \lambda$ with zero temperatures of protons and electrons. For the sake of simplicity, we form a symmetric electron vortex, artificially generating the localized magnetic field distribution during a number of initial timesteps. As we are not interested in the physical reasons of vortex formation, but would like to discuss the evolution of these coherent structures, this method is suitable for our purposes.
Thus, we add the following profile of the magnetic field in our numerical scheme
\begin{equation}
    B_{\rm add}(x,y) = B_0 \cdot \exp\left( -\left(\frac{x - x_0}{l_{\rm vort}}\right)^2 - \left(\frac{y - y_0}{l_{\rm vort}}\right)^2 \right),
\end{equation}
\noindent where $B_0$ is the maximum amplitude of the added magnetic field profile, $l_{\rm vort}$ is the typical vortex width, $x_0$ and $y_0$ are initial coordinates of the vortex center. We use this addition to the magnetic field calculations each timestep during the first 5 time units. For $B_0 = 10^{-3}$ and $l_{\rm vort} = 0.5 \lambda$, at time equals to 10, we observe a quasistatic vortex structure with the maximum magnetic field amplitude of $0.4$ and $6 \lambda$ in diameter. These parameters are chosen in the way to satisfy the conditions for ``magnetized electrons and unmagnetized ions'' approach that is used in the theoretical description of the electron vortex evolution \cite{Gordeev}. Quantitavely speaking, we set the value of the maximum magnetic field inside the cavity satisfying the relation 
\begin{equation}
    \left( \frac{\omega_{pe}}{\omega_0} \right)^2 \ll a_B ^2 \ll \left( \frac{\omega_{pe}}{\omega_0} \right)^2 \cdot \left(\frac{m_p}{m_e} \right),
\end{equation} which means that the field energy is enough to expel the electrons out from the axis region of the vortex (left inequality), but not strong enough to influence proton dynamics significantly \cite{Gordeev}. The computational grid is $40 \lambda \times 40 \lambda$ with 32 nodes per each cell. The initial particle-in-cell number is equal to 16. The total number of particles is about $2 \times 10^7$. The integration timestep equals to 0.0125 time units. The total time of simulations is 300 time units.

\section{Evolution of the electron vortex}

\begin{figure}
    {a)}\includegraphics[width=0.45\linewidth]{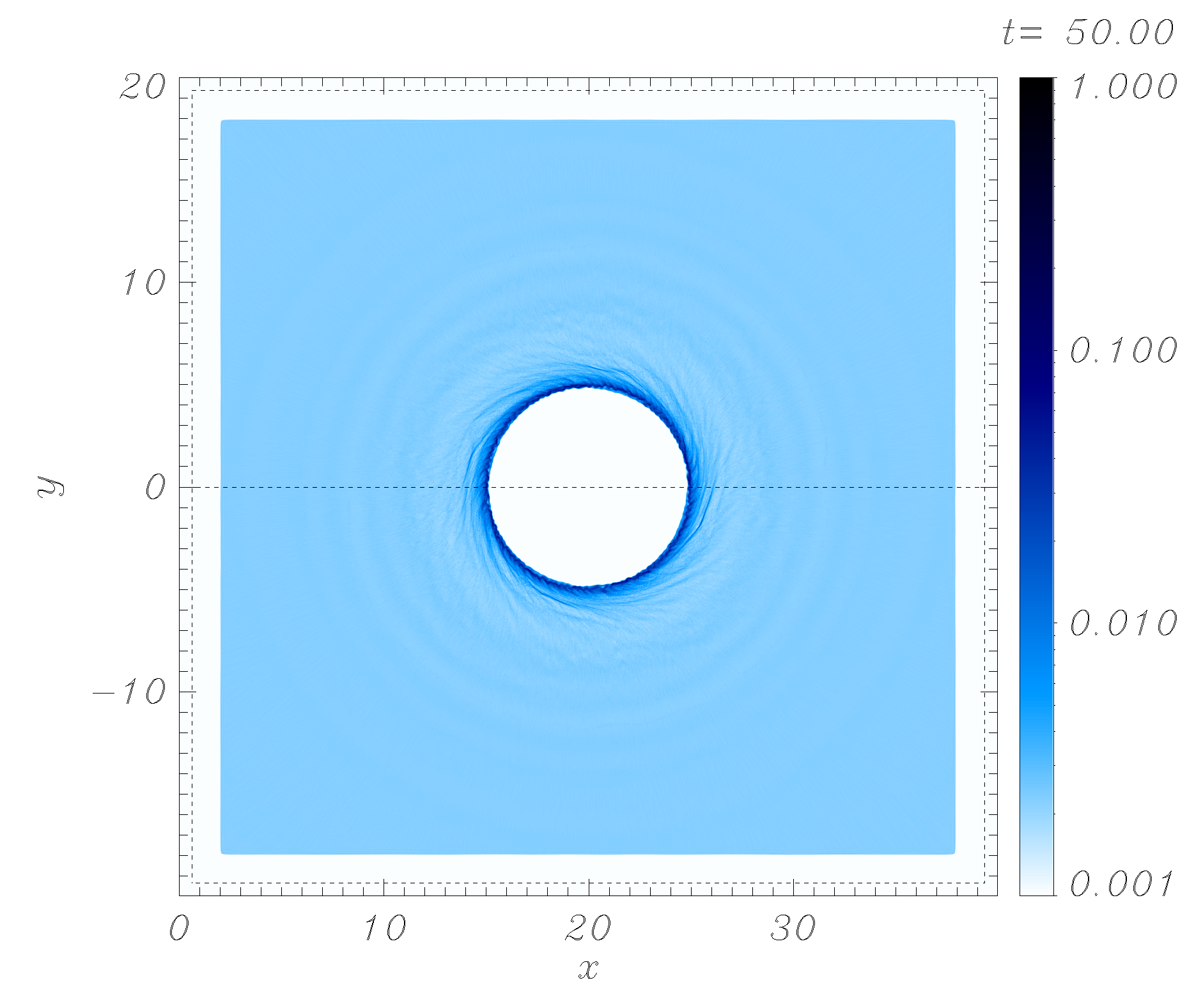}
    {b)}\includegraphics[width=0.45\linewidth]{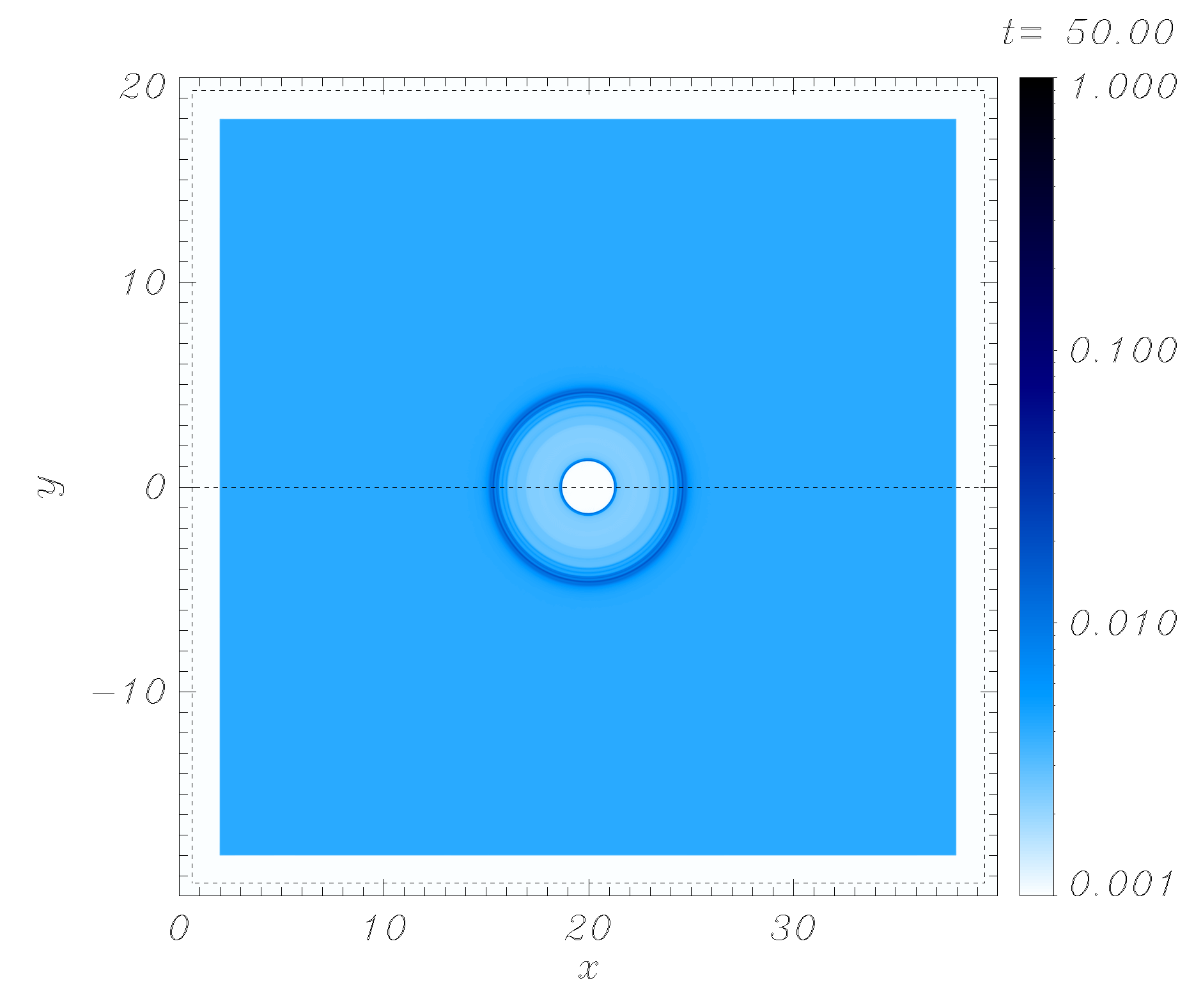}\par
    {c)}\includegraphics[width=0.45\linewidth]{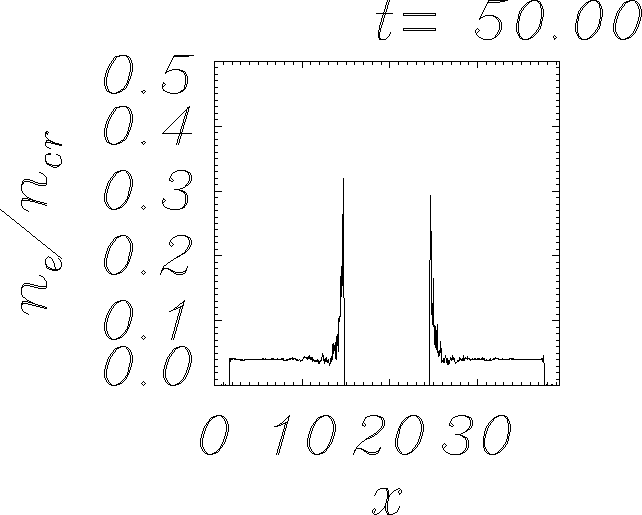}
    {d)}\includegraphics[width=0.45\linewidth]{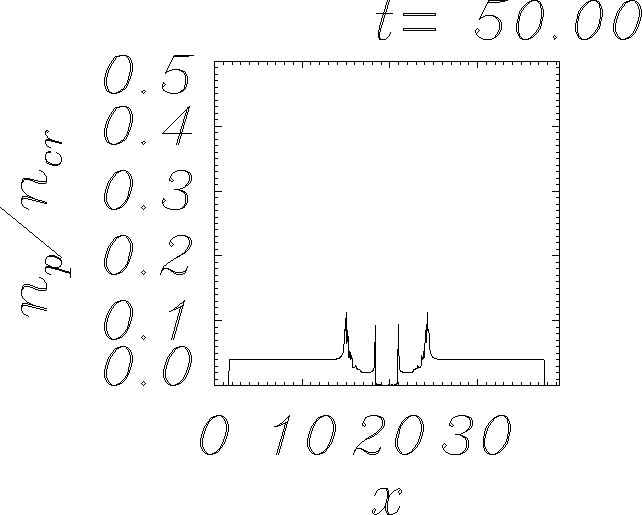}\par
    {e)}\includegraphics[width=0.45\linewidth]{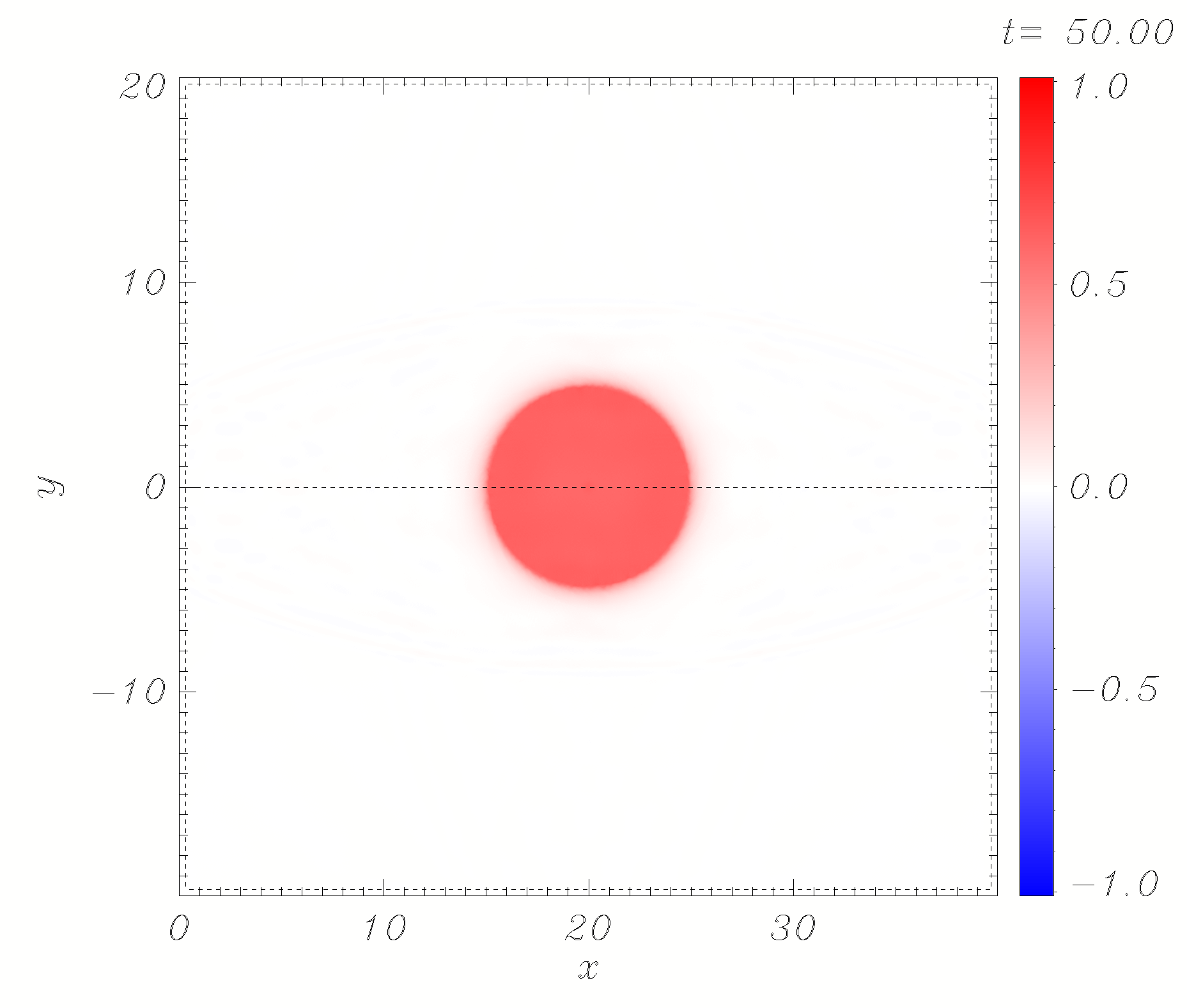}
    {f)}\includegraphics[width=0.45\linewidth]{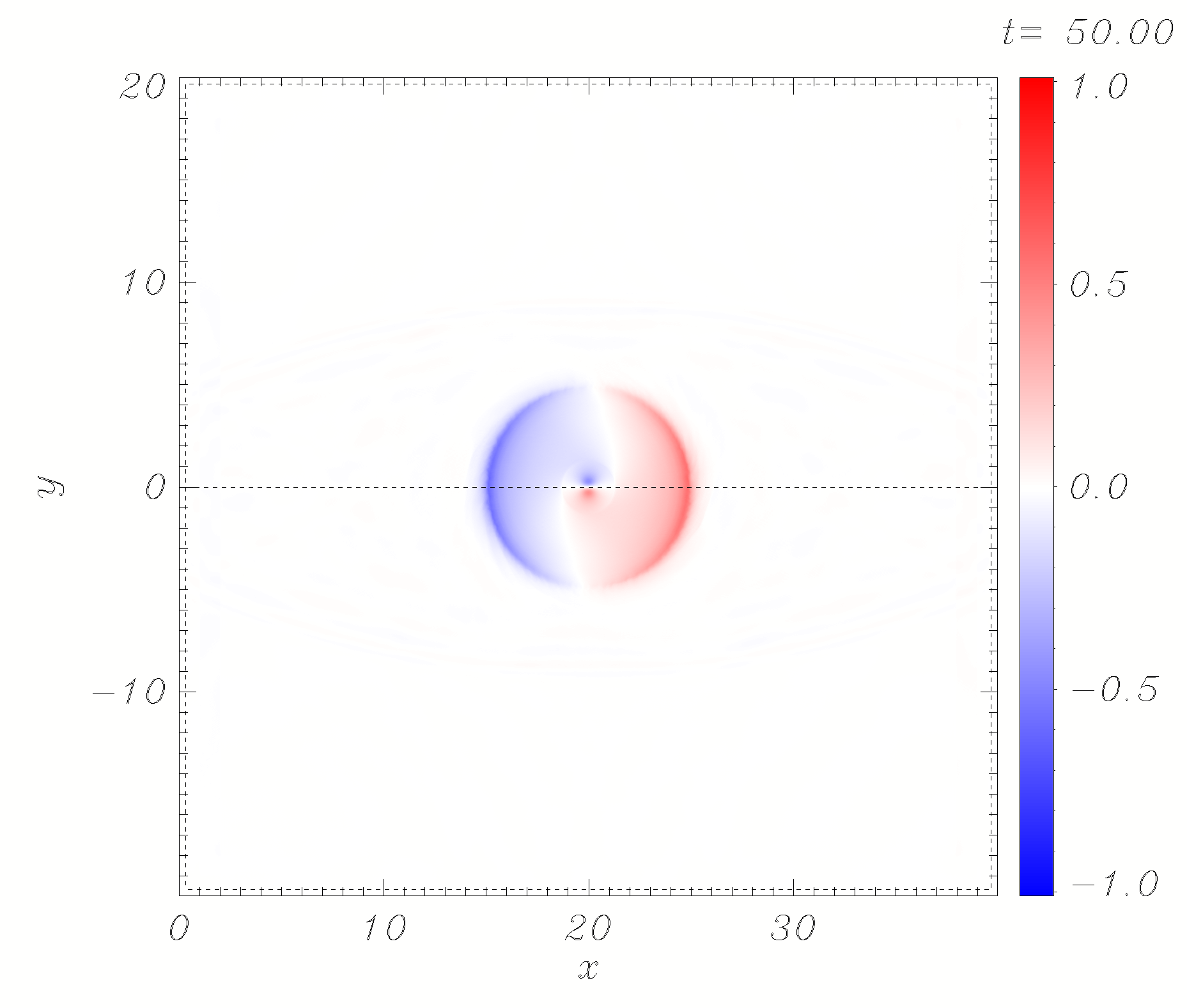}\par
\caption{a) electron, b) proton density distributions, c) electron and d) proton density profiles along $y=0$, e) $B_z$ component of the magnetic field and f) $E_x$ component of the electric field for t=50.}
\label{fig:firstdens}
\end{figure}

\begin{figure}
    \includegraphics[width=\linewidth]{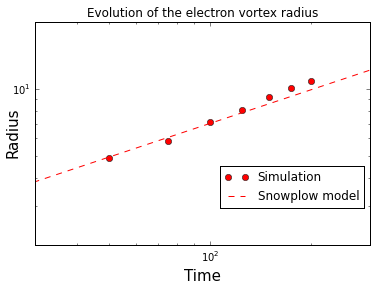}
    \caption{Evolution of the vortex radius over time: simulation data versus snowplow model. We observe quite a good agreement, with the best fit $R \sim t^{0.58}$, in comparison to the power law from snowplow model $R \sim t^{0.5}$.}
    \label{fig:asympt}
\end{figure}

As we expected, the simulation with immobile ions justifies that there is a quasistationary configuration of electron fluid rotating in the localized magnetic field, which is similar to one described in Section II. However, on a proton timescale $\sim\omega_{\rm pi}^{-1}$, we find out that these vortices expand due to noncompensated positive charge enclosed in the electron cavity. Here we consider how the vortex evolves during the simulation time. In case of $t=50$ (see Figure \ref{fig:firstdens}) we observe an axisymmetric electron cavity with a substantinal increase of the electron density around the shock-like vortex boundary, up to $0.4 n_{\rm cr}$. 
The magnetic flux is concentrated inside the cavity. These facts give us a description of the evolution of the vortex using the snowplow model \cite{Sakai}. Let us assume that the magnetic flux inside the cavity is constant in time (i.e., we assume an adiabatic approximation). Then, we can suppose that the whole dynamics is determined by the magnetic field pressure on a walls of axisymmetric vortex. We write

\begin{equation}
    \frac{\rm d}{\rm dt} \left( M \frac{{\rm d}R}{\rm dt} \right) = 2 \pi R \frac{ B^2}{8 \pi}, 
\end{equation}

\noindent where $M=\pi n_0 m_i R^2$ is the mass of plasma pushed to the radius $R$ by the magnetic flux and $\langle \bold B^2 \rangle = \langle \bold B_{in}^2 \rangle \times (R_{in}/R)^2$, where index $in$ denotes to the initial values of the magnetic field and vortex radius. We can rewrite this equation in dimentionless variables $r=R/R_{in}$ and $\tau = \sqrt{B_{in}^2 / 4 \pi m_i n_0}~t$:

\begin{equation}
    \frac{\rm d}{\rm d\tau} \left(r^2 \frac{{\rm d}r}{\rm d \tau} \right) = \frac{1}{r},
\end{equation}

\noindent which is the same as in \cite{Sakai}. The solution could be written as follows:

\begin{equation}
    \frac{1}{2}(r \sqrt{r^2-1}+\ln{\sqrt{r^2-1}+r})=\tau-\tau_{in},
\end{equation}

\noindent for which the asymptotics in $t \rightarrow \infty$ limit are: $R\sim t^{1/2}$, $\dot{R} \sim t^{-1/2}$, and $B \sim t^{-1/2}$  (note the misprint in \cite{Sakai}). Comparing the obtained asymptotics with simulations, we plot the Figure \ref{fig:asympt}. We observe quite a good agreement of the radius evolution between theory and simulations, despite the simplification of theoretical model in comparison with the simulation setup.

\begin{figure}
    {a)}\includegraphics[width=0.46\linewidth]{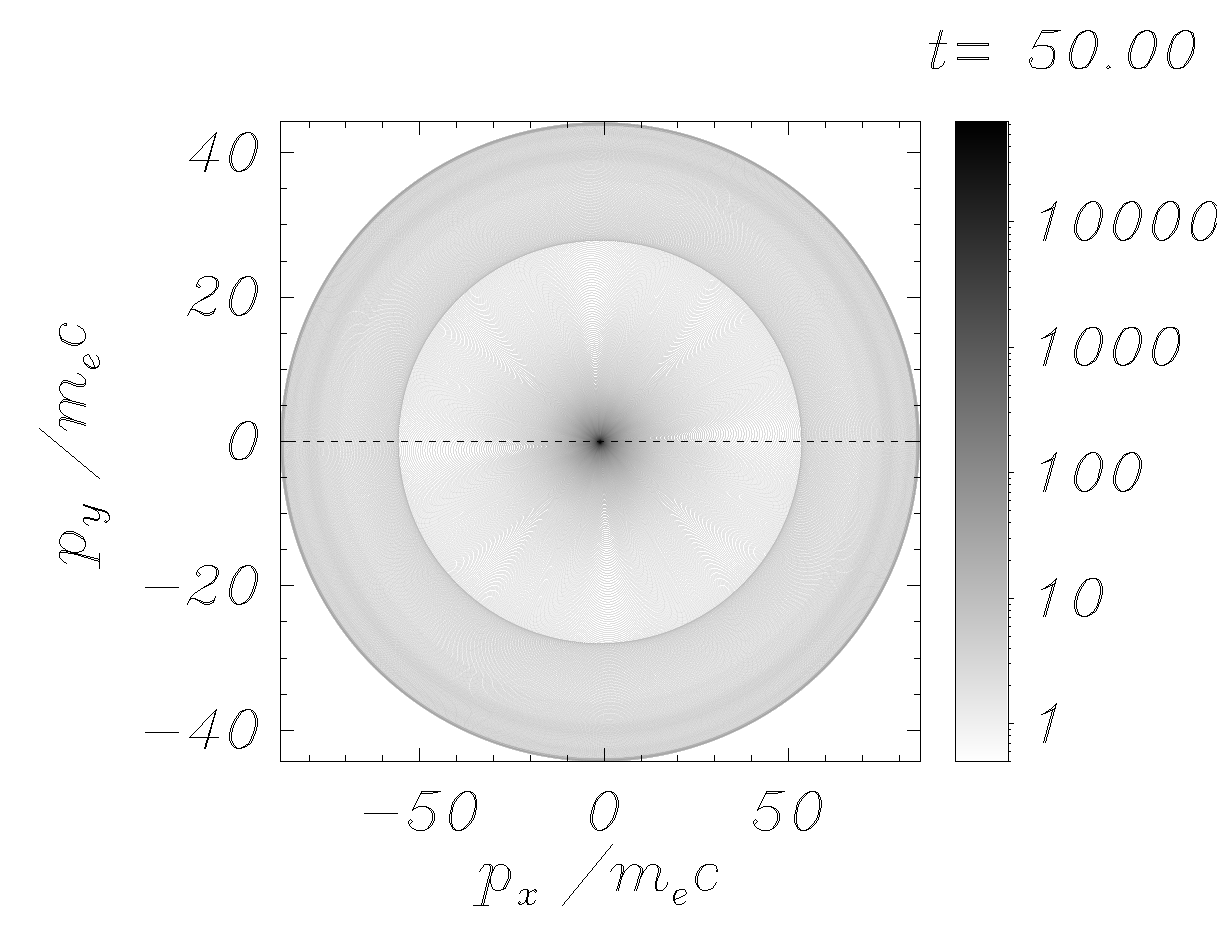}
    {b)}\includegraphics[width=0.46\linewidth]{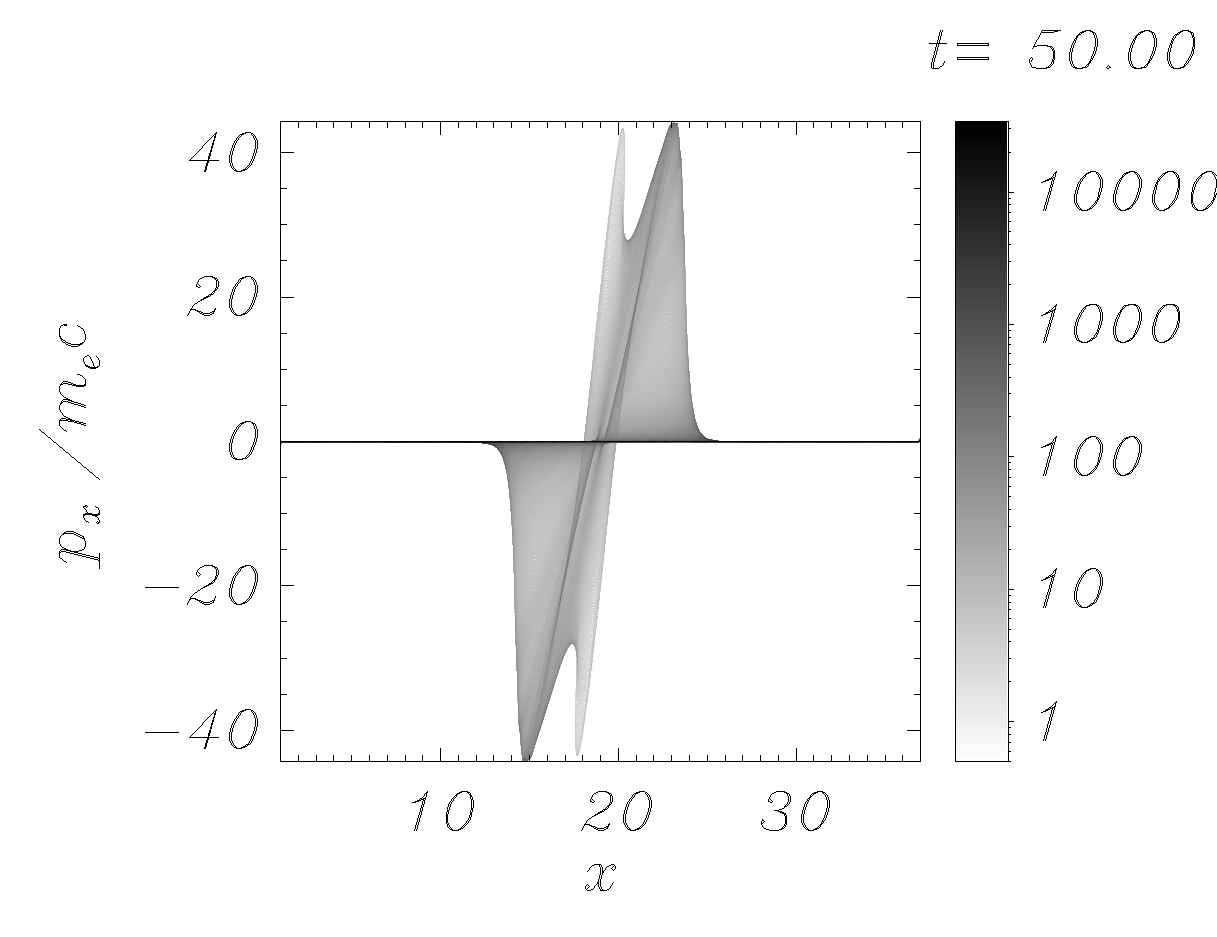}
\caption{Proton phase space distributions: a) $x-p_x$ and b) $p_x - p_y$ for t=50.}
\label{fig:firstphase}
\end{figure}

It is also interesting to discuss the proton phase space distributions. Figure \ref{fig:firstphase} shows the $x - p_x$ and $p_x - p_y$ phase space distributions of protons. At $x-p_x$ phase plot we may see two antisymmetric regions, corresponding to Coloumb explosion in the external plasma -- the region where $p_x \sim x$ correspond to the Coloumb explosion of the cyllindrical noncompensated positive charge of protons. The subsequent decrease of the proton momentum is due to the finite velocity of interacting electromagnetic field propagation. 
While exploding due to the Coloumb repulsion, more and more protons are involved in the process, starting to move radially. 
The $p_x-p_y$ figure shows that there are numerous proton shells that move radially in an axisymmetric way. The shell-nature of the dynamics leads to a number of multistream instabilities that come into play during the simulation.

\begin{figure}
    {a)}\includegraphics[width=0.45\linewidth]{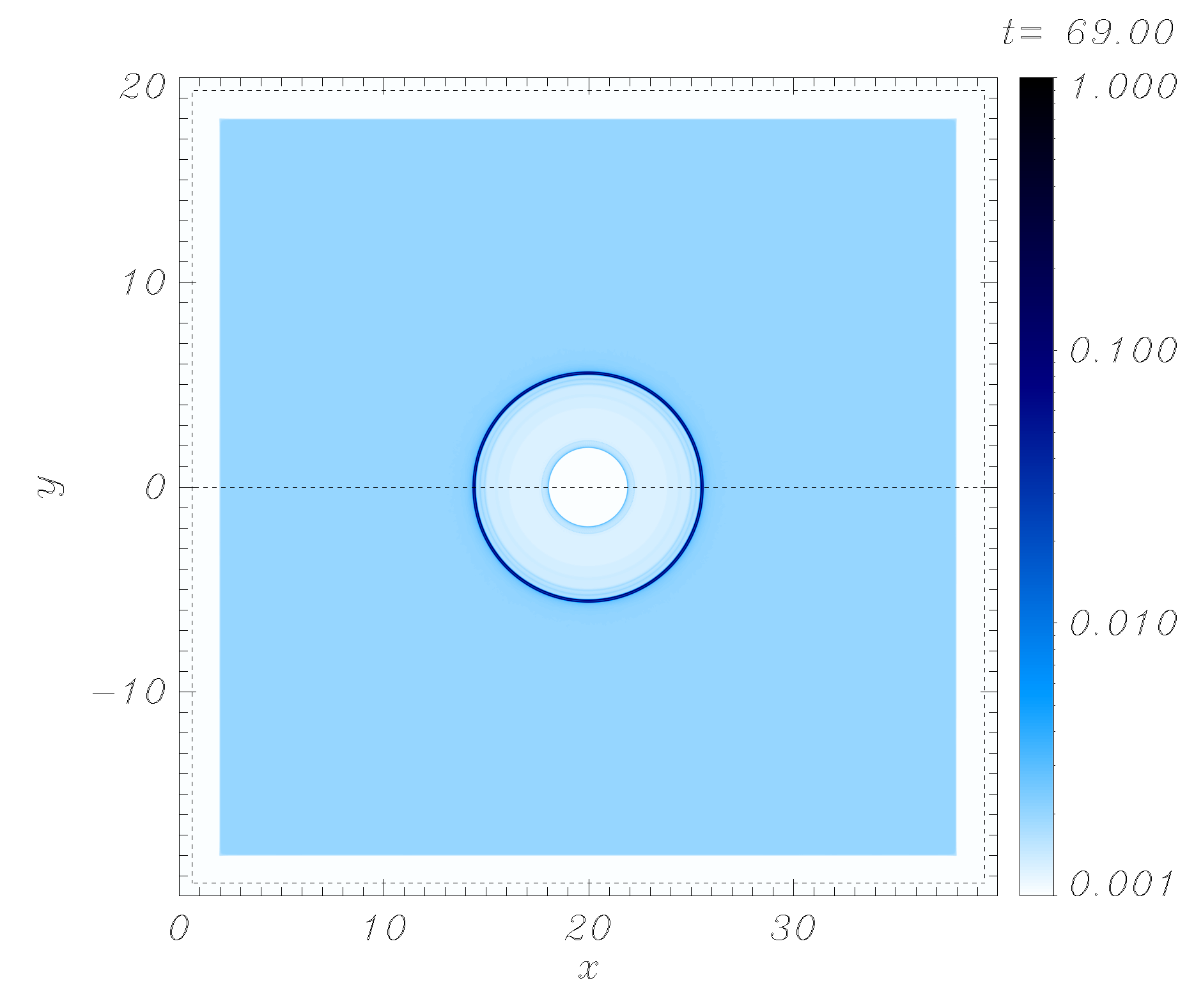} 
    {b)}\includegraphics[width=0.45\linewidth]{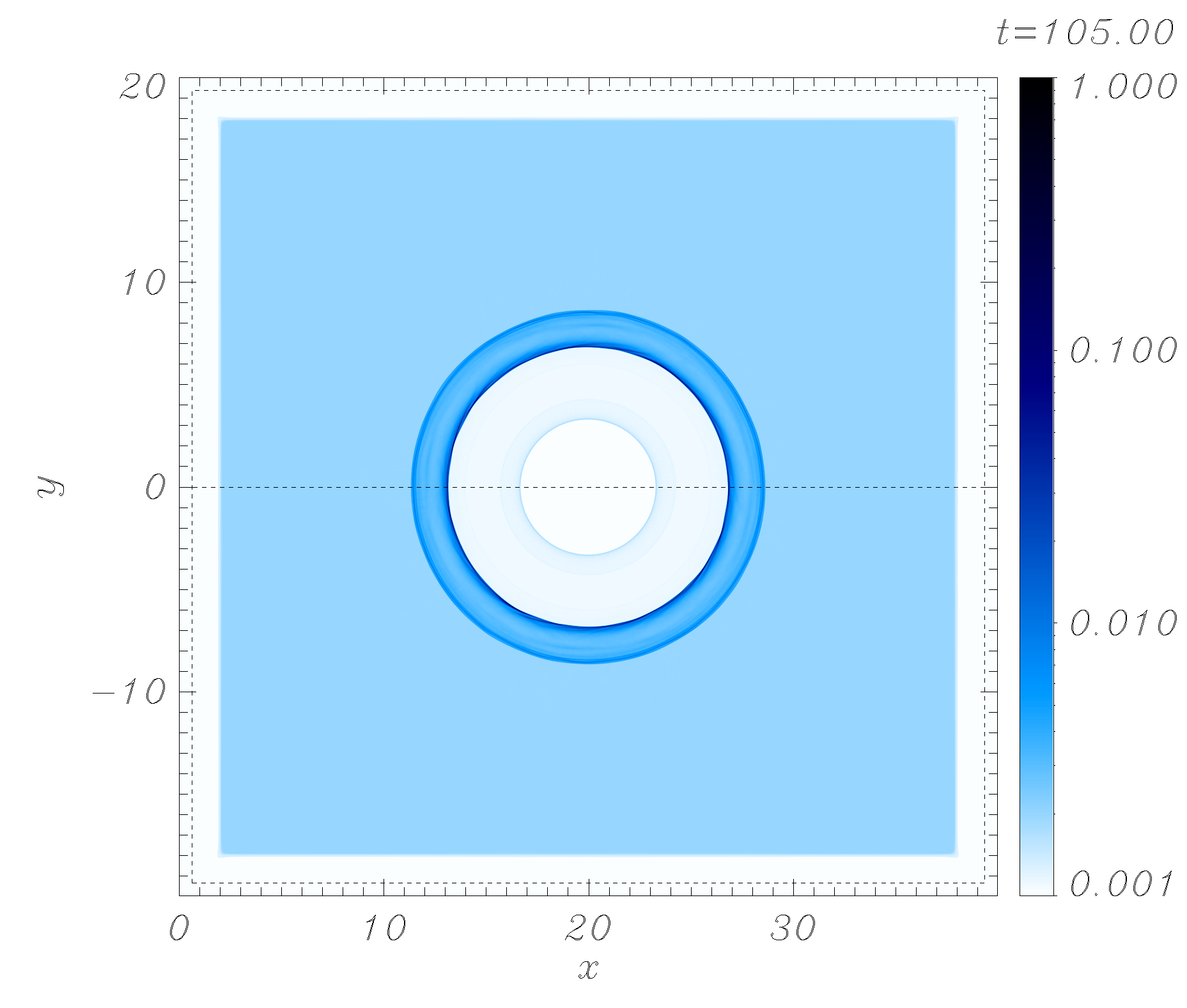}\par
    {c)}\includegraphics[width=0.45\linewidth]{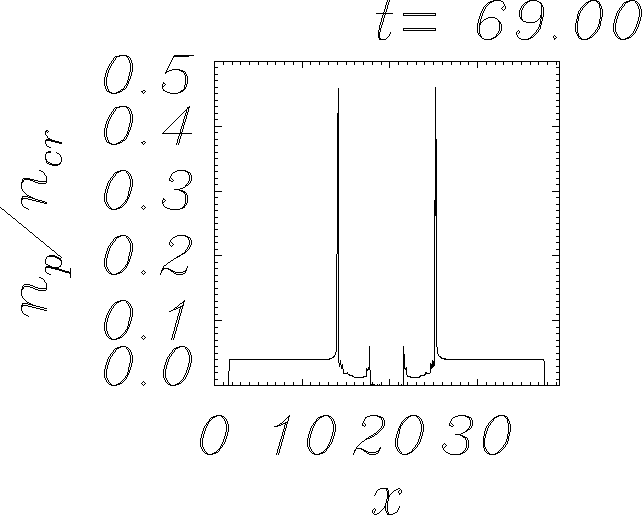}
    {d)}\includegraphics[width=0.45\linewidth]{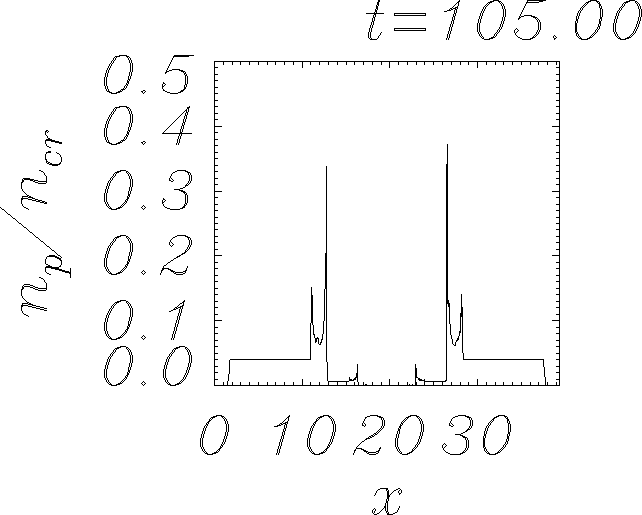}\par
    {e)}\includegraphics[width=0.45\linewidth]{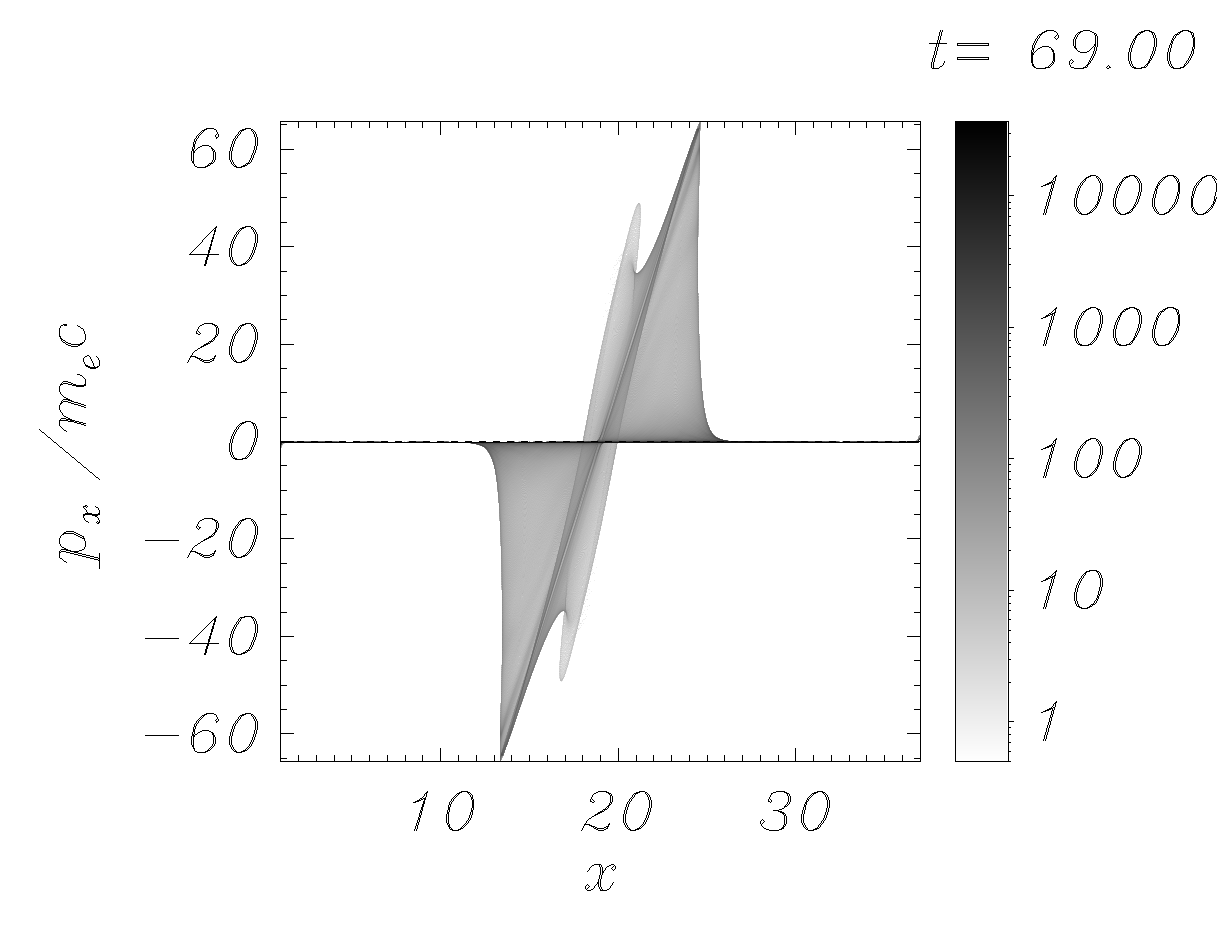}
    {f)}\includegraphics[width=0.45\linewidth]{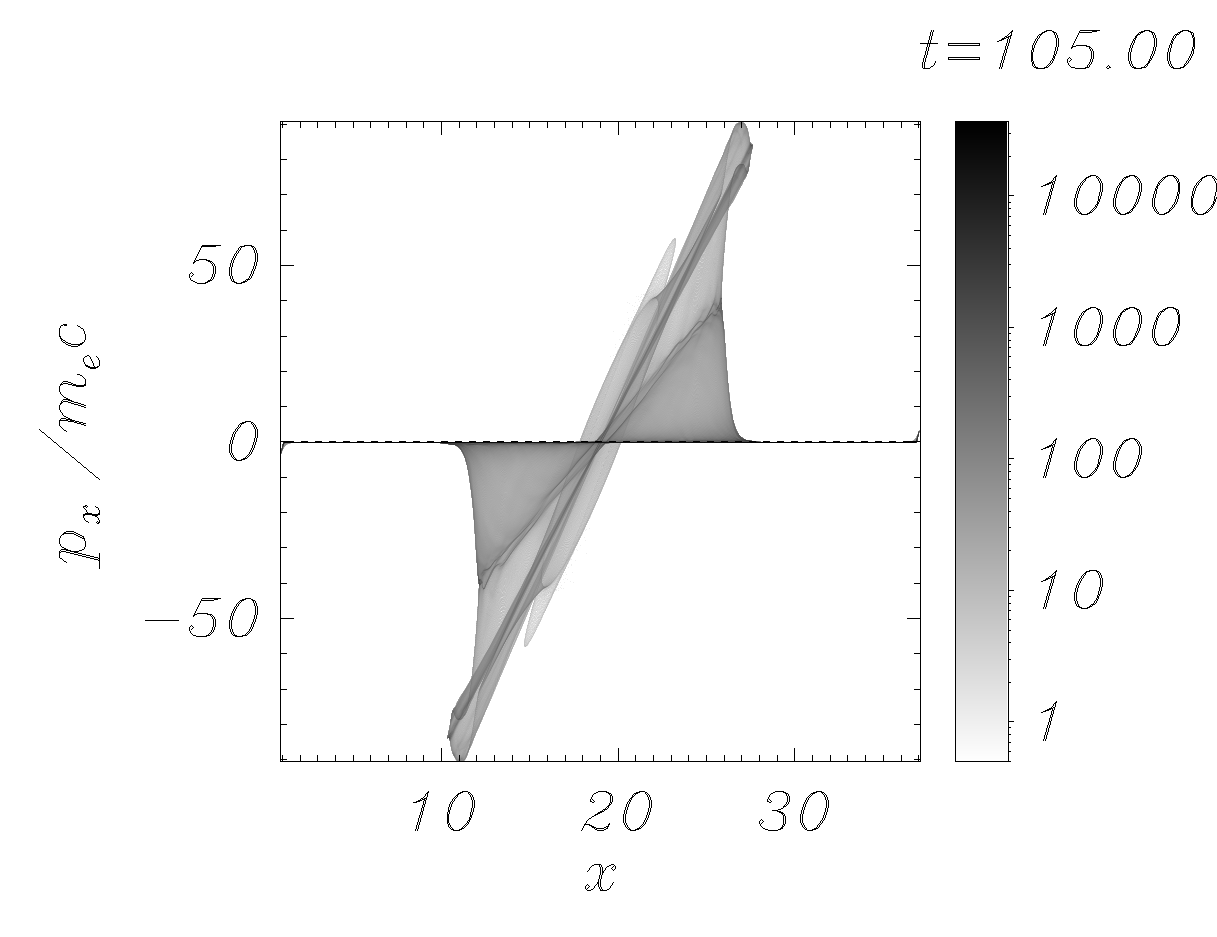}\par
    \par
    {g)}\includegraphics[width=0.49\linewidth]{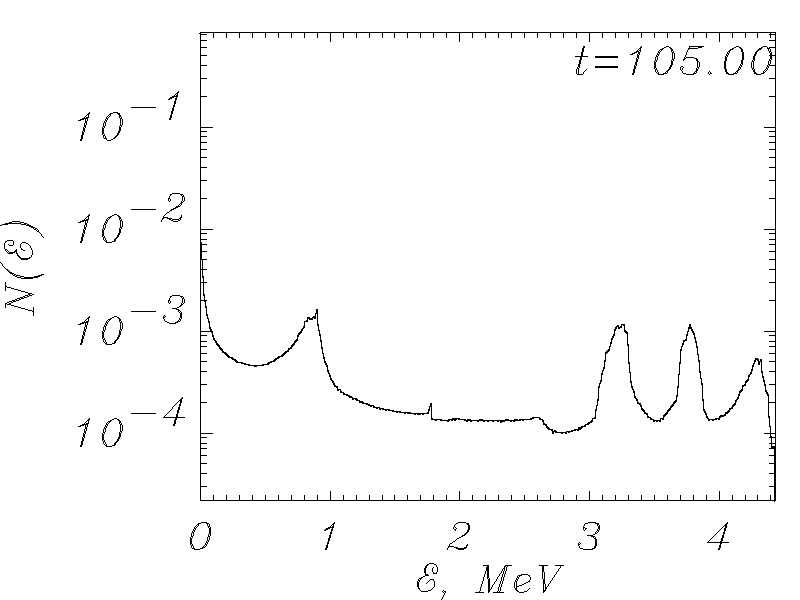}
\caption{Instability manifestation: a) proton density (c) - proton density profile along $y=0$), e) $x - p_x$ phase plot for protons - correspond to the moment of $\rm d p_x /dx \rightarrow \infty$ at $t=69$, b), d), and f) - the same for the moment of manifestation of the multistream instability, $t=105$, g) proton energy spectrum at $t=105$, three peaks corresponding to the most energetic ion associations are seen.}
\label{fig:instability}
\end{figure}

At Figure \ref{fig:instability}, we observe that the $\rm d p_x / dx$ derivative tends to infinity in $x-p_x$ phase plots. Physically, it means that there are multiple proton shells located at the same spatial coordinate but having the broad range of energies. This reveals a rise of a multistream instability, leading to the enhanced proton energy gain rate.
We plot the same figures for $t=105$. Here, we see that the dense proton ring has been divided into the three most energetic parts, which are seen at both density profile and energy spectrum of protons. 
It is also noticeable that there are some phase space folds revealing on a $x-p_x$ plot, that also correspond to these independently moving proton circles.

\begin{figure}
    {a)}\includegraphics[width=0.46\linewidth]{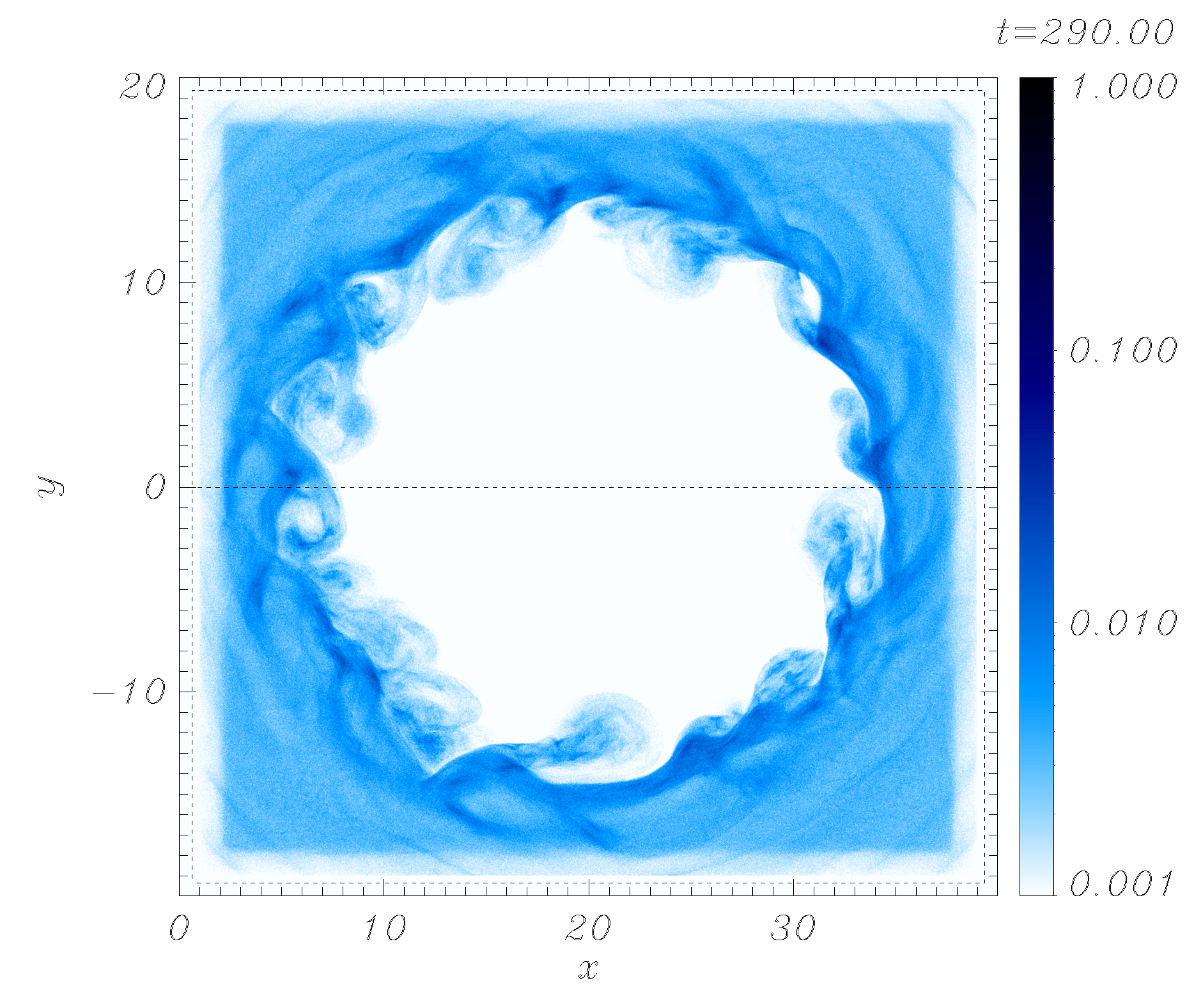} 
    {b)}\includegraphics[width=0.46\linewidth]{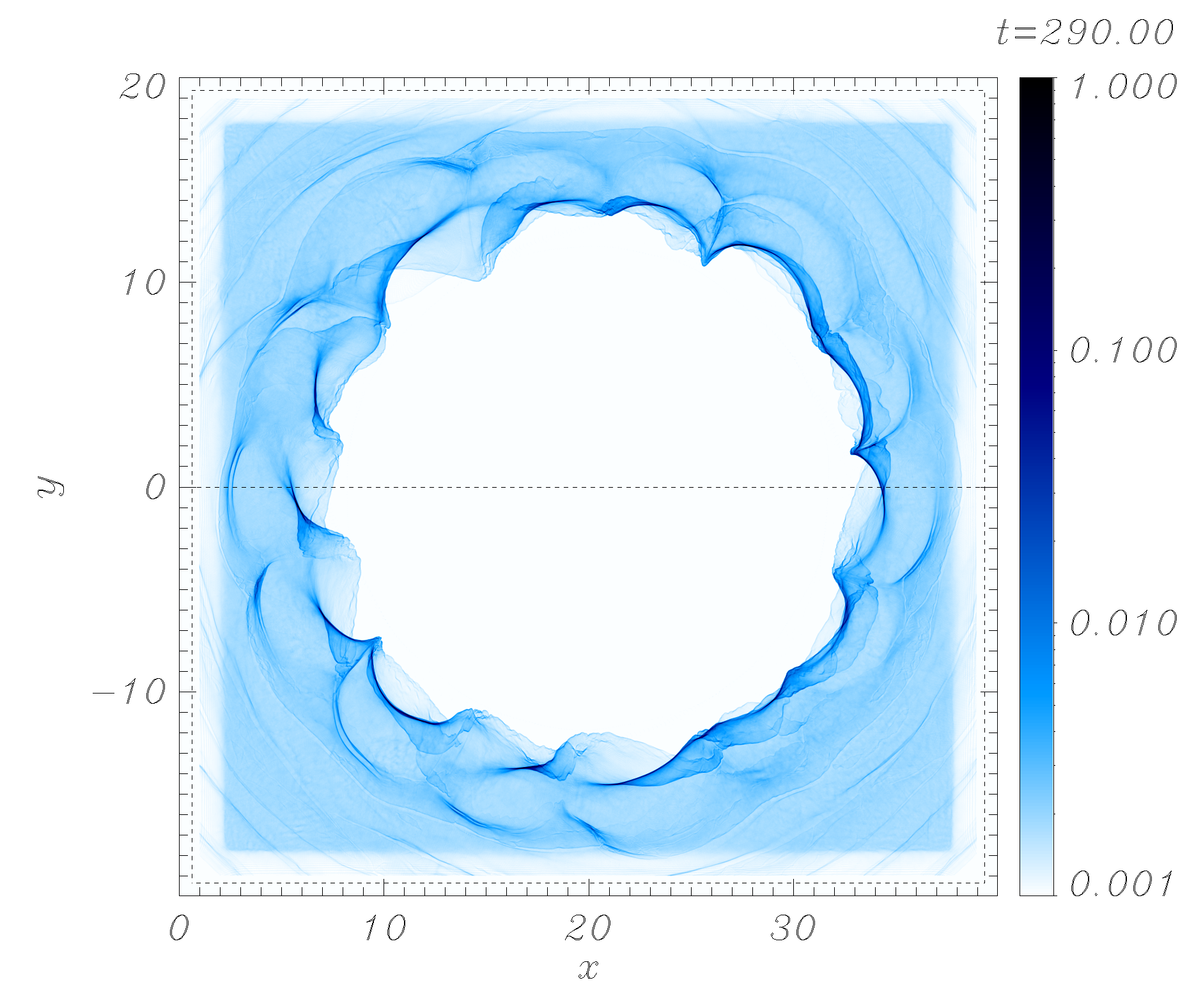}
    \par 
    {c)}\includegraphics[width=0.46\linewidth]{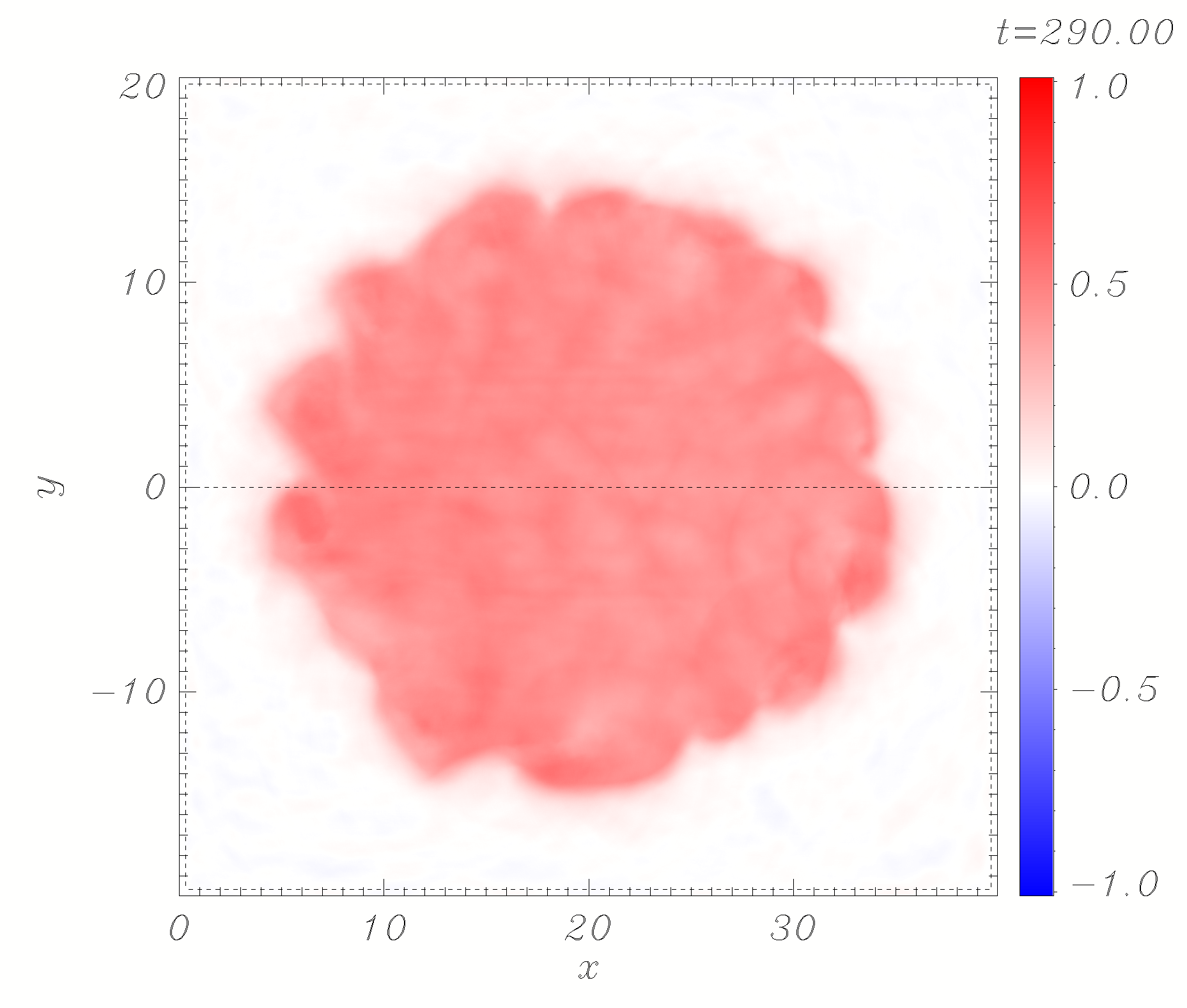}
    {d)}\includegraphics[width=0.46\linewidth]{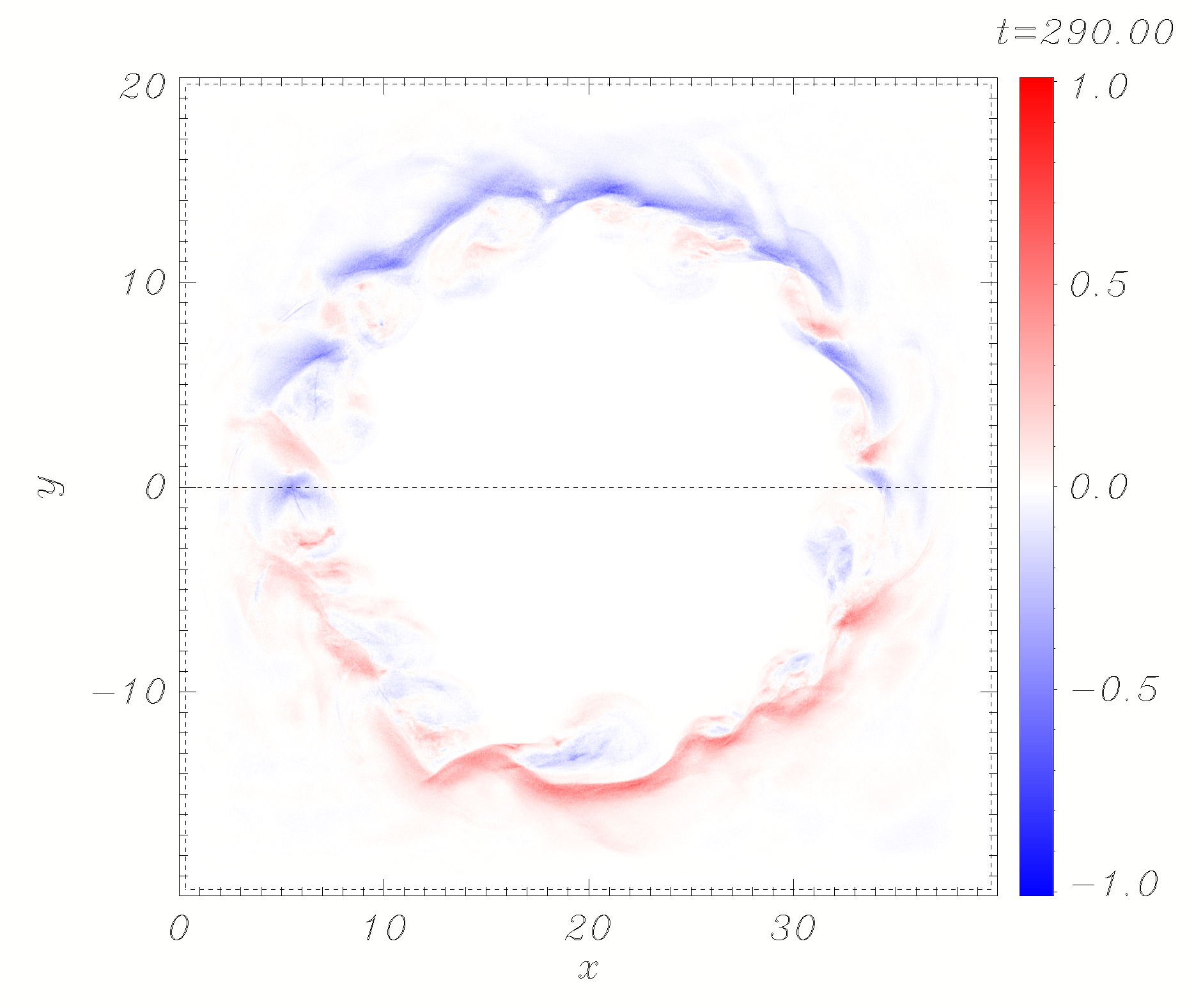}
    \par
    {e)}\includegraphics[width=0.49\linewidth]{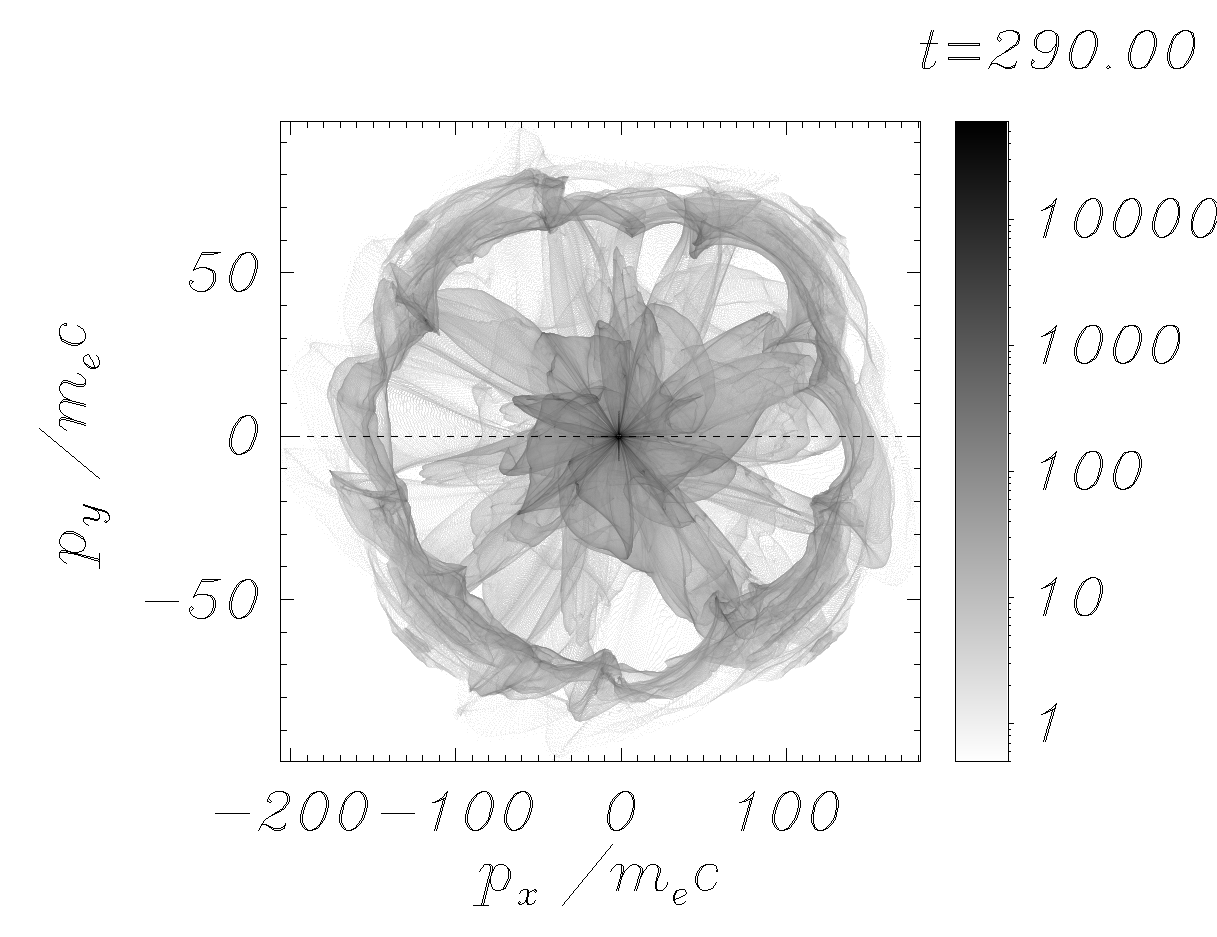}
    \caption{Small vortices formation: a) electron, b) proton densities, c) z-component of magnetic field $B_z$, d) x-component of electric current $C_x$, and e) proton phase plot in $p_x - p_y$ coordinates for $t=290$. A group of small vortices along the vortex boundary are seen.}
\label{fig:vortices}
\end{figure}

After some time of vortex evolution, we observe the effect of bending of the electric current associated with the electron vortex. Besides, the vortex boundary starts to break into a group of small electron vortices. Figure \ref{fig:vortices} shows electron and proton densities, $z$-component of the magnetic field and $x$-component of electric current. This effect may correspond to the manifestation of various instabilities, having some similarity with the hydrodynamical instability of vortex boundary \cite{VORTEXRES} and the electromagnetic Kelvin-Helmholtz instability \cite{PICKH}. It is also important to note that not only electron dynamics is affected by this instability, as we observe the anisotropy in proton acceleration as well, see Figure \ref{fig:vortices}e).

\section{Conclusion}

In this article, we presented analytical description and computer simulation results on the stationary state and evolution of one type of coherent structures that are observed in laser plasmas - electron vortices. These structures are often seen in 2D PIC simulations of various laser-plasma configurations. While being quasistatic in the immobile ion approach, they do evolve on the ion timescale for the two-species simulations. We observed the manifestation of various stream instabilities, which enhance maximum energies of protons and lead to shell-like structure of proton distribution. We have obtained $5 \rm ~MeV$ protons accelerated by the combination of Coloumb explosion of the mildly relativistic vortex with $B_{\rm max} \sim 10^4$ T $\times(1~ \mu{\rm m}/\lambda)$ and multistream instability.




At final stages of the vortex evolution, ion rotational motion becomes significant in addition to their expansion. This leads to disintegration of the vortex boundary into the association of a small electron vortices, making an impact on a ion acceleration as well. The obtained results will be useful for developing the theory describing the electromagnetic turbulence in relativistic plasmas \cite{TURBPLASMA1, TURBPLASMA2} and diagnostics of the results of the experiments with petawatt-class laser systems, that are being build nowadays \cite{ELI-BL}.

\section{Acknowledgements}

This work was supported by the ELI Project No. CZ.02.1.01/0.0/0.0/15\_008/0000162. We also would like to acknowledge the support from Russian Foundation for Basic Research (grant No. 15-02-03063). 

\end{document}